\documentclass[twocolumn]{iopart}
\usepackage{iopams}
\usepackage{graphicx}
\usepackage{dcolumn}

\begin{document}

\title[Electronic structure of the superconductor UCoGe]
{Electronic structure of the ferromagnetic superconductor UCoGe from first principles}

\author{Pablo de la Mora$^{1,2}$ and O Navarro$^3$}
\address{$^1$ Departamento de F\'isica, Facultad de Ciencias,
Universidad Nacional Aut\'{o}noma de M\'{e}xico, 
Apartado Postal 70-542, 04510 M\'{e}xico D.F., M\'{e}xico}
\address{$^2$ Instituto de Investigaciones Metal\'urgicas, Universidad 
Michoacana de San Nicol\'as de Hidalgo, Morelia, Michoac\'an, M\'exico}
\ead{delamora@servidor.unam.mx}
\address{$^3$ Instituto de Investigaciones en Materiales,
Universidad Nacional Aut\'{o}noma de M\'{e}xico,
Apartado Postal 70-360, 04510 M\'{e}xico D.F., M\'{e}xico.}


\begin{abstract}
The superconductor UCoGe is analyzed with electronic structure calculations 
using Linearized Augmented Plane Wave method based on Density Functional 
Theory. Ferromagnetic and antiferromagnetic calculations with and without 
correlations (via LDA+U$_H$) were done. In this compound the Fermi level
is situated in a region where the main contribution to DOS comes from
the U-5f orbital. The magnetic moment is mainly due to the Co-3d orbital
with a small contribution from the U-5f orbital. The possibility of fully
non-collinear magnetism in this compound seems to be ruled out. These 
results are compared with the isostructural compound URhGe, in this case
the magnetism comes mostly from the U-5f orbital.
\end{abstract}


\section{Introduction}

Considerable theoretical and experimental effort has been focused on the
interplay of magnetism and superconductivity since it is widely believed
that it could help to clarify the mechanism behind high-$T_c$
superconductivity \cite{khb}. Particular interest has been put on to the 
coexistence of
ferromagnetism and superconductivity because of the nontrivial phenomena
which are predicted or found experimentally, such as the still controversial
triplet pairing symmetry found in ferromagnetic superconductors.

After the BCS theory of superconductivity \cite{bcs,buzdin}, it became clear
that pairing of electrons in the singlet state could be destroyed by an
exchange mechanism, such as the exchange field in a magnetically ordered
state who tends to align spins of Cooper pairs in the same direction
preventing a pairing effect. However, experimental evidence has been 
found of the coexistence of antiferromagnetic order and superconductivity 
in ternary rare-earth compounds \cite{maple82}. Superconductivity and 
antiferromagnetism can coexist, on average, because the exchange and orbital
fields are zero at distances of the order of the Cooper pairs size. Later 
on it was demonstrated that ferromagnetic order is unlikely to appear in 
the superconducting phase \cite{anderson59, buzdin}. In spite of
this, ferromagnetism was found in superconductors and the 
first evidence of ferromagnetic superconductors was given in bulk materials; 
UGe$_2$ (at high pressure) \cite{saxena} and URhGe (at ambient pressure) 
\cite{aoki}, it has been 
argued that critical magnetic fluctuations could mediate superconductivity 
in these compounds.

These systems seem to have triplet paring symmetry which permits the 
coexistence of superconductivity
and ferromagnetism. In this direction, Huy \textit{et al.} \cite{Huy} recently
report, superconductivity on the weak ferromagnet UCoGe at ambient pressure.
They claim that superconductivity ($T_c$=0.8K) and ferromagnetic order ($T
_{C}$=3K) indeed coexist. They also found an agreement with the triplet
pairing scenario in the UCoGe compound. The magnetic moment found by Huy 
\textit{et al.} for this compound is M=0.03$\mu _{B}$, which contrasts with 
the effective 
paramagnetic moment M=1.7$\mu _{B}$ \cite{Troc}. They claim that this material
should have band magnetism.

In this paper, spin polarized electronic structure analysis was performed to 
further 
understand the UCoGe superconductor. The objective of this analysis is to 
correlate the particular characteristics of this material with its
superconducting properties.

\section{Computational details}

Electron quantum mechanical calculations were done with the WIEN2k package 
\cite{Blaha}, which is a Linearized Augmented Plane Wave (FP-LAPW) method
based on Density Functional Theory (DFT). Spin-orbit coupling is included in
a second-variational way, and the strong correlations in uranium 
were included via LDA+U$_H$(SIC) (contributed by Pavel Nov\'{a}k). 
The Generalized Gradient Approximation of
Perdew, Burke and Ernzerhof \cite{Perdew} was used for the treatment of the
exchange-correlation interactions. The energy threshold to separate
localized and non-localized electronic states was -6 Ry. The muffin-tin
radii were: 2.5$a_0$ for uranium, 2.43$a_0$ for cobalt, 2.48$a_0$ for rhodium, 
and 2.15$a_0$ for germanium ($a_0$ is the Bohr radius). The criterion for the 
number of plane waves was $R_{MT}^{min}\times K^{max}=9$ and the number of 
k-points were 462 ($7\times 11\times 6$), for the case with all the atoms 
independent it was 225 ($5\times 9\times 5$). For crystal structure 
visualization the XCrySDen package \cite{Kolkalj} was used.

\begin{figure}[t]
\begin{indented}
\item[]\includegraphics[width=8cm,height=7.5cm]{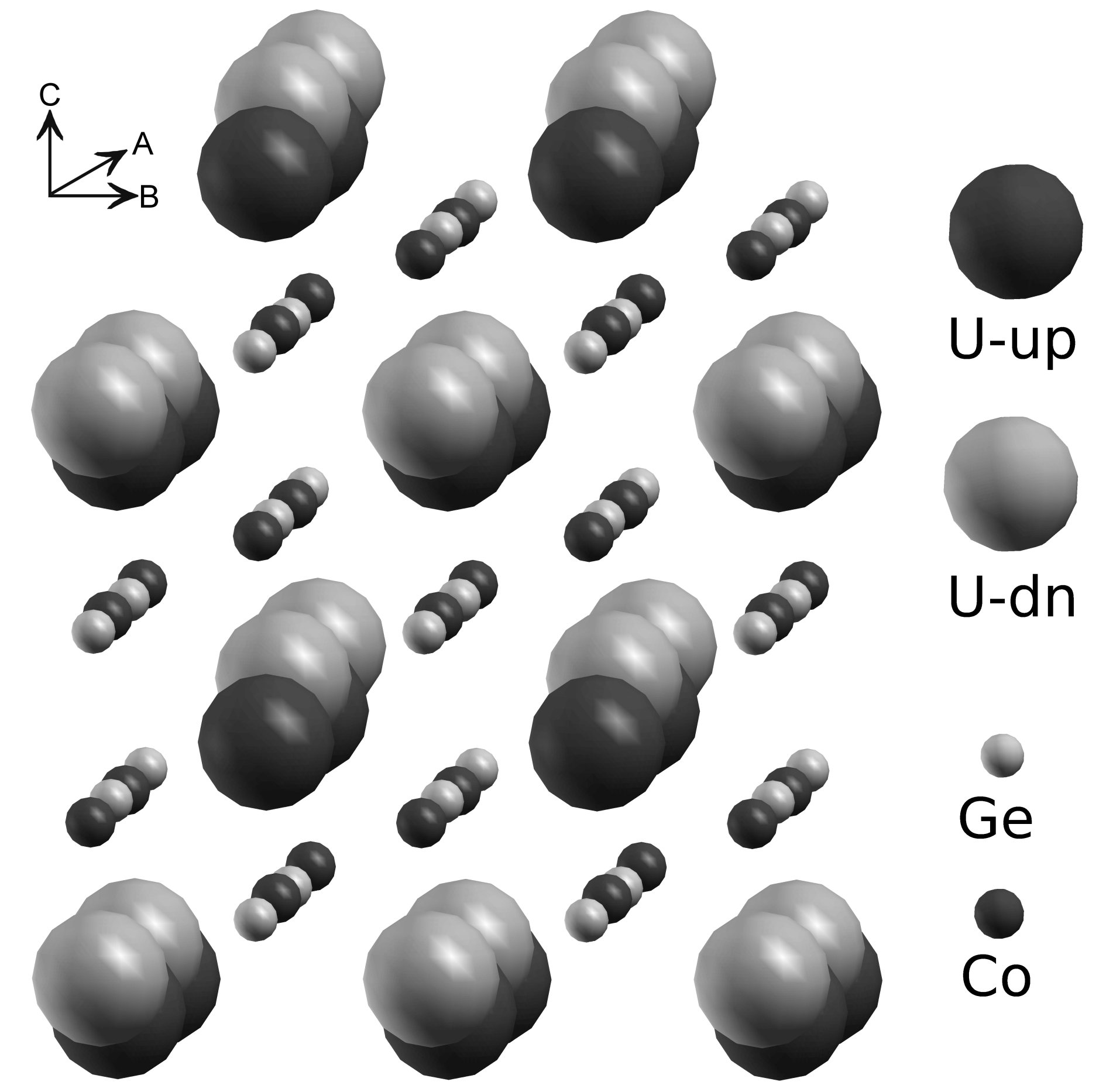}
\caption{\label{figura1}{\protect\small {Crystal structure of UCoGe.\\ 
U-up and U-dn refer to the antiferromagnetic cell.}}}
\end{indented}
\end{figure}

\section{Results and discussion}


\begin{table}
\caption{\label{tabla1} Atom positions in the unit cell. 
The unit cell is orthorhombic 
with space group $Pnma$ (SG \#62), the cell parameters are 
$a$=6.845\AA, $b$=4.206\AA, $c$=7.222\AA.}
\begin{indented}
\item[]\begin{tabular}{@{}cccc}
\br
& $a$ & $b$ & $c$ \\
\mr
U & 0.0101 & 0.25 &  0.7075 \\ 
Co & 0.2887 & 0.25 & 0.4172 \\ 
Ge & 0.1967 & 0.25 & 0.0870 \\
\br
\end{tabular}
\end{indented}
\end{table}

UCoGe has the same crystal structure as URhGe \cite{Canepa, URhGe-cita}, 
it belongs to the
space group $Pnma$ (SG \#62). The crystal structure is shown in
Figure 1, with the cell parameters taken from Huy \textit{et al.} 
\cite{Huy}, and the internal atomic positions from Canepa \textit{et al.} 
\cite{Canepa}. These parameters are shown in Table 1.

The crystal structure is very similar to that of the MgB$_2$ compound, 
which consists of B-graphene sheets intercalated with Mg atoms. In the 
case of UCoGe, it is formed of Co-Ge-graphene sheets intercalated with
U atoms, as shown in Figure 1 (in the same arrangement as LiBC 
\cite{Rosner}). The U atoms form chains perpendicular to the grephene
sheets, but contrary to MgB$_2$ and LiBC these chains are not straight but 
they are in a zigzag arrangement (see Figure 1), and the graphene sheets become
corrugated, that is, the Co-Ge-hexagons that are between two U atoms are 
almost in the plane perpendicular to the U-U line.

Of the U-atoms substructure each U atom has four U neighbours, 
two along the chain, and other two in contiguous chains. The four 
neighbours form an unsymmetrical tetrahedron around the central U atom. 
This situation is similar to C in diamond, that is, the U atoms form a 
distorted diamond structure.

For electronic structure calculations where there are heavy elements it is 
important to include spin-orbit interactions, 
and the magnetization is now dependent of the crystallographic direction. 
The results show that the most stable magnetization direction is in the 
$c$-axis by 
3.65meV and 3.00meV with respect to the $a$- and $b$-directions 
respectively, which is the same direction as in URhGe \cite{Divis, Shick}.

\begin{table}
\caption{\label{tabla2} Magnetic moments ($\mu_B$) in the three different 
directions (only the main orbital component is shown), the last column 
corresponds to URhGe in the $c$-direction.}
\begin{indented}
\item[]\begin{tabular}{@{}ccccccc}
\br
\textbf{Direction} & & $a$ & $b$ & $c$ & & $c$ (Rh) \\ 
\mr
\textbf{Spin term} & U & 1.088 & 1.090 &  1.083 & & 1.037\\ 
& Co (Rh) & -0.487 & -0.482 & -0.472 & & -0.108\\ 
& Ge & -0.028 & -0.028 & -0.026 & & -0.026\\ 
\mr
\textbf{Orbital term} & U & -0.961 & -1.105 &  -1.181 & & -1.266\\ 
& Co (Rh) & -0.058 & -0.064 & -0.063 & & -0.011\\ 
\mr
\textbf{Sum} & U & 0.128 & -0.016 &  -0.098 & & -0.228\\ 
& Co (Rh) & -0.546 & -0.546 & -0.534 & & -0.119\\
\mr
\textbf{Total} & & -0.418 & -0.562 &  -0.633 & & -0.348 \\
\br
\end{tabular}
\end{indented}
\end{table}

The magnetic moments are shown in Table 2, the spin components vary little 
with the crystal direction, but the orbital moments vary considerably. In the
$c$-direction the total moment of the U atom is very small 
(M=0.098$\mu _{B}$). The 
situation of the Co atom is quite different; the orbital moment is quite small 
but the spin component is fairly large and the total moment does not cancel 
(M=0.633$\mu _{B}$), these values are close to those found by  Divi\v{s} 
\cite{Divis1}, although he found a smaller total moment; M=0.28$\mu _{B}$. 
This result is in disagreement with the experimental results of Huy 
\textit{et al.} \cite{Huy}, in which the total magnetic moment is quite 
small (M=0.03$\mu _{B}$), but it is still smaller than the effective 
paramagnetic 
moment found by Tro\'{c} and Tran (M=$1.7\mu _{B}$) \cite{Troc}.
These results contrast with the case of URhGe, where the Rh atom has 
a small spin moment and a smaller orbital moment; it is the U atom that has 
the largest contribution.

\begin{figure}[t]
\begin{indented}
\item[]\includegraphics[width=7cm,height=11cm]{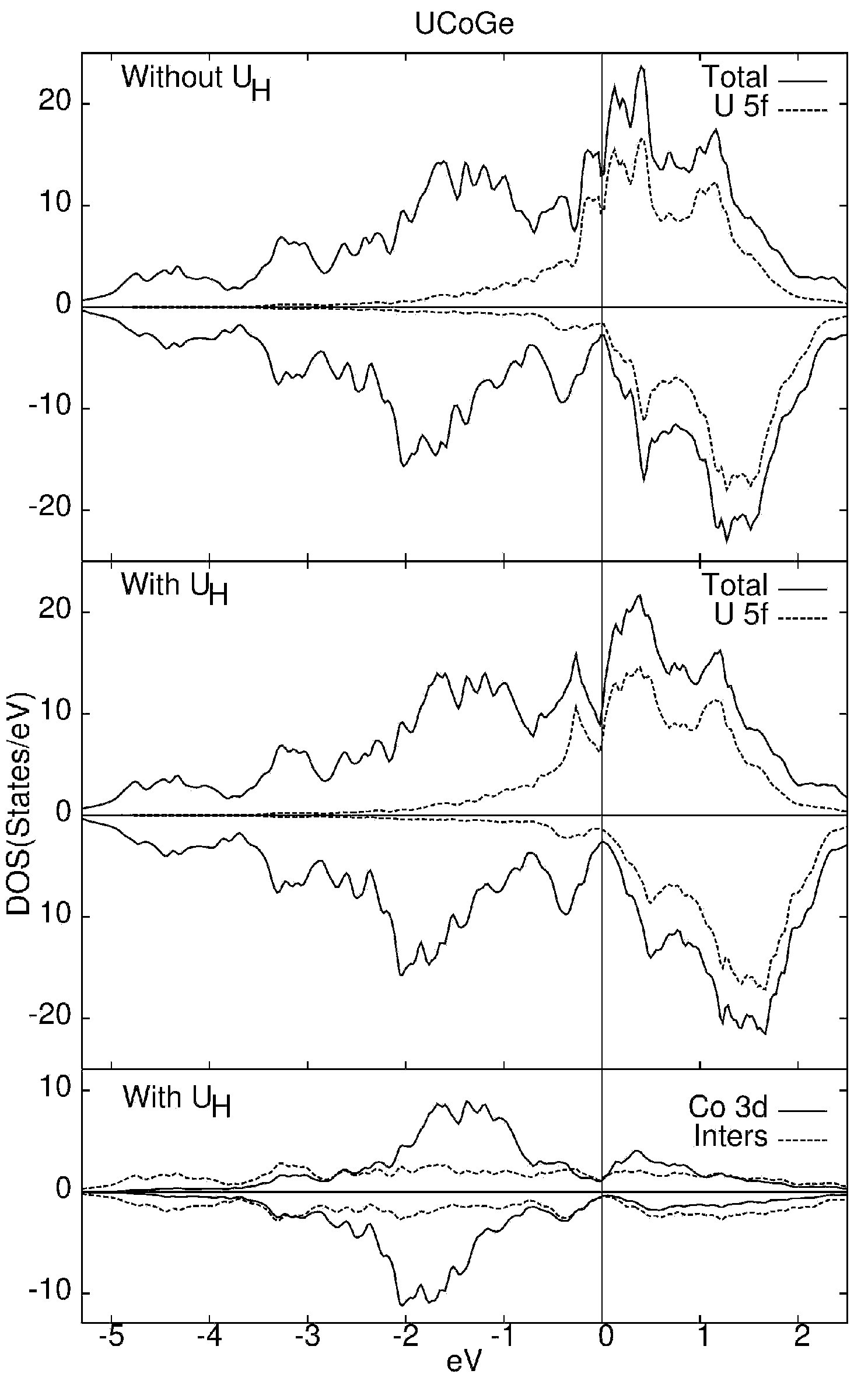}
\caption{\label{Figure2}{\protect\small {DOS for UCoGe ferromagnetic calculation, 
up-spin: upward, dn-spin: downward. Top: without $U_H$ correction, middle 
and bottom with $U_H$ correction.
The main contribution at the $E_F$ is from U, at -1.5eV
cobalt has a large contribution. Ge has very little contribution everywhere 
and is not shown.}}}
\end{indented}
\end{figure}

The U-5f orbital is fairly localized and there is a strong intra-coulomb 
repulsion energy, this effect is not normally included in the DFT calculations,
but it can be included via a $Hubbard-U_H$ term ($LDA+U_H(SIC)$) in the WIEN2k
package. The meaning of the $U_H$ term was discussed by Anisimov and 
Gunnarsson \cite{Anisimov} who defined it as the cost in Coulomb energy by
placing two electrons in the same site. They also devised a method of 
calculating the $U_H$ term from first principles using a supercell. 
Madsen and Nov\'ak \cite{Madsen} adapted this method to the FP-LAPW method 
(WIEN2k). 
Using this method the effective $U_H$ ($U^{eff}_H=U_H-J$), for the U-atoms in
UCoGe, was found
 to be $U_H$=0.362eV. Rusz and Divi\v{s} for UPtAl found that 
$U^{eff}_H$=0.36eV \cite{Rusz} fit the experimental values best. For the other 
atoms no $U_H$ was used.


\begin{table}
\caption{\label{tabla3} Magnetic moments ($\mu_B$) of UCoGe and URhGe in the 
$c$-direction with $U_H$-correction (only the $c$-direction is shown)}
\begin{indented}
\item[]\begin{tabular}{@{}ccccc}
\br
\textbf{Compound} & & UCoGe & URhGe \\ 
\mr
\textbf{Spin term} & U & 1.175 & 1.151 \\ 
& Co/Rh & -0.524 & -0.122 \\ 
& Ge & -0.031 & -0.029\\ 
\mr
\textbf{Orbital term} & U & -1.341 & -1.470\\ 
& Co/Rh & -0.076 & -0.015\\ 
\mr
\textbf{Sum} & U & -0.167 & -0.319\\ 
& Co/Rh & -0.600 & -0.137\\
\mr
\textbf{Total} & & -0.767 & -0.456\\
\br
\end{tabular}
\end{indented}
\end{table}

With the $LDA+U_H$ correction the magnetic values increase, see Table 3 
(for URhGe the same $U_H$ value was used). For UCoGe with this increment
the total magnetic moment (M=0.767$\mu _{B}$) moves further away from the 
experimental value of M=0.03$\mu _{B}$. 
On the other hand, the total magnetic moment for 
URhGe increases to M=0.456$\mu _B$, which is quite close to the 
experimental value reported by Aoki \textit{et al.} (M=0.42$\mu_B$)
\cite{aoki}. For this compound the experimental values vary quite considerably
from M=0.19$\mu _B$ \cite{Tran} to a value about three times larger 
\cite{Boer}. As mentioned above for 
UCoGe, if both the magnetic (M=0.03$\mu _{B}$) and the paramagnetic
(M=1.7$\mu _{B}$) values are analyzed, 
the situation is more drastic, and further experimental measurements 
would narrow this large difference.

Due to the discrepancy between the calculated and small experimental magnetic
value in UCoGe, an antiferromagnetic calculation was performed. 
This is in view of a possible scenario that would explain the small 
experimental total magnetic moment; in an antiferromagnetic configuration 
the individual magnetic moments are antiparallel and the total magnetic 
moment is zero. If these moments are not exactly antiparallel but slightly
canted then they would not cancel exactly and a small perpendicular 
moment would remain, this would explain the small experimental value.

The U atoms that form a distorted diamond structure can be separated into 
U-up and U-dn now with a distorted zincblende structure, the new crystal 
structure belongs to the space group $Pnm2_1$ (SG \#31). 
The collinear antiferromagnetic configuration with the magnetic moment 
in the $c$-direction was calculated. The energy was found to be 33.2meV
higher than that of the ferromagnetic configuration. The canted configuration
should not be very different from the collinear one in terms of energy 
(the angle would be of $\sim$4.3$^\circ$), 
therefore the canted antiferromagnetic configuration can be ruled out,
and the ferromagnetic configuration with the magnetization in the 
$c$-direction is the most stable configuration.

\begin{figure}[t]
\begin{indented}
\item[]\includegraphics[width=7cm,height=11.4cm]{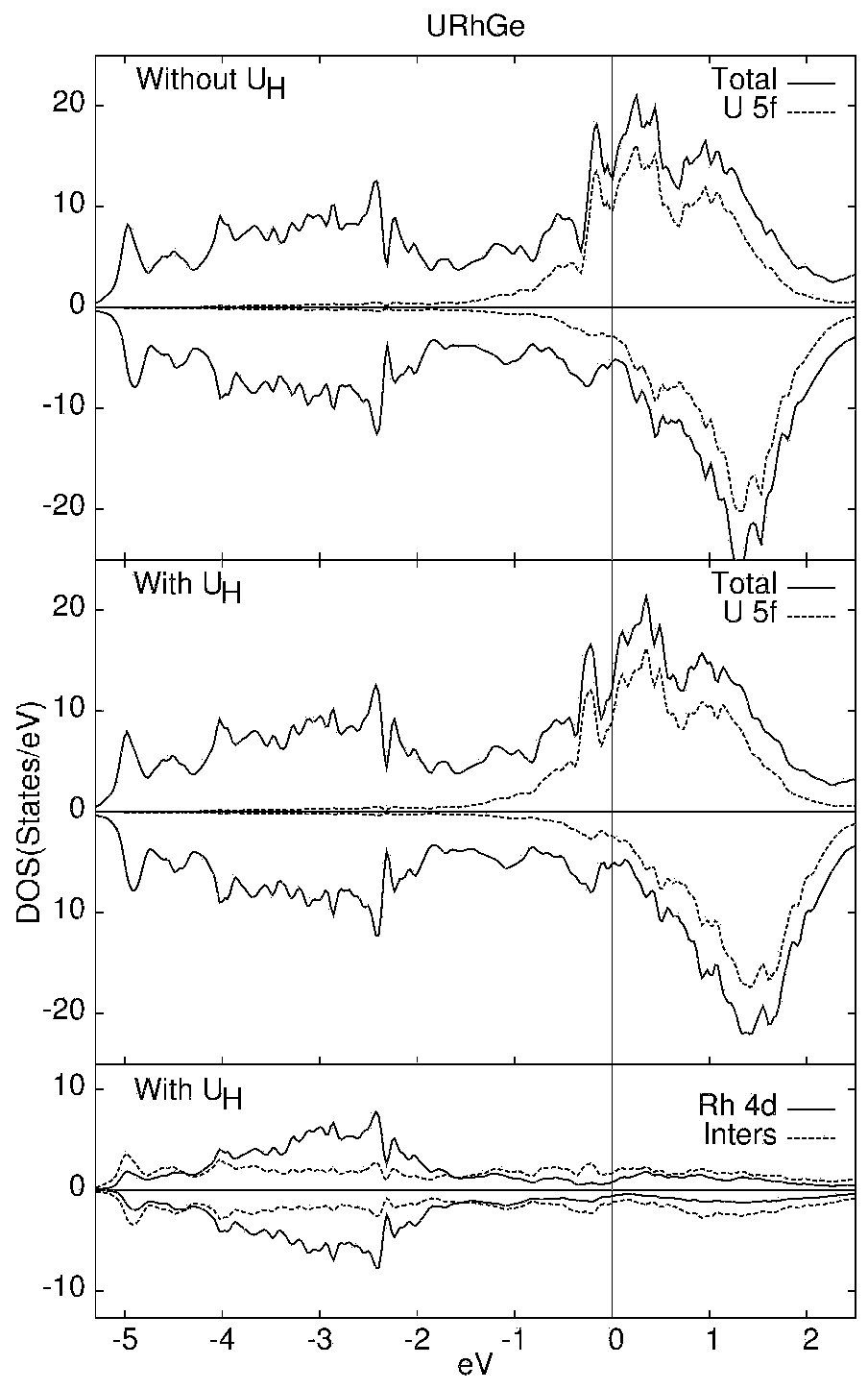}
\caption{\label{Figure3}{\protect\small {DOS for URhGe ferromagnetic 
calculation,
rhodium has a small contribution at $E_F$.}}}
\end{indented}
\end{figure}


Electronic structure calculation for the ferromagnetic structure with
the magnetic moments in the $c$-direction (Figure 1) shows that all the
atomic bonds are mostly metallic, that is, the charge density is around the
atoms and has an almost spherical shape, in the interstitial space the
density is fairly homogeneous. The main contribution to the 
\textit{Density of States} (DOS) (Figure 2) comes from U that is mostly U-5f
followed by Co that is mostly Co-3d. There is a large U peak at $E_F$ 
[-0.8eV, 2eV], Co has a large contribution at $\sim$ -1.5eV, but still close
enough to $E_F$ and Co has an important magnetic moment. The Ge 
contribution, not shown, is very small, there is an important contribution 
from the 
interstitial space. DOS at $E_F$ is quite large for up-spin and it is small 
for down-spin, when the $U_H$-correction is introduced a small gap begins to 
form for up-spin. DOS for URhGe, Figure 3, shows that the Rh-4d is at much 
lower energy, $\sim$ -3eV, and the magnetic moment is considerably smaller 
than in UCoGe.

\begin{table}
\caption{\label{tabla4} Magnetic moments ($\mu_B$) of UCoGe with the principal 
magnetization in the $a$-direction, with all the 
atoms independent, the canting is in the $c$-direction}
\begin{indented}
\item[]\begin{tabular}{@{}cccccccccc}
\br
\textbf{Component} & \textbf{Spin} &  & \textbf{Orbital} & \\ 
Canting         &  & (c) & & (c) \\ 
\mr
\textbf{U} & 1.133 & -0.032 & -0.991 &  0.141 \\ 
           & 1.224 &  0.028 & -1.183 & -0.208 \\ 
           & 1.223 & -0.029 & -1.183 &  0.210 \\ 
           & 1.133 &  0.032 & -0.994 & -0.140 \\ 
\mr
\textbf{Co} & -0.498 &  0.001 & -0.044 & -0.005 \\ 
            & -0.534 &  0.005 & -0.124 & -0.009 \\ 
            & -0.533 & -0.005 & -0.124 &  0.009 \\
            & -0.499 & -0.001 & -0.044 &  0.005 \\
\br
\end{tabular}
\end{indented}
\end{table}

\begin{table}
\caption{\label{tabla5} Magnetic moments ($\mu_B$) of UCoGe with all the 
atoms independent, the canting direction is in parenthesis}
\begin{indented}
\item[]\begin{tabular}{@{}cccccccccc}
\br
\textbf{Direction} & \textbf{a} &  & & \textbf{b} & & \textbf{c} & \\ 
Canting         &  & (c) & &  & &  & (a) \\ 
\mr
\textbf{U} & 0.142 &  0.109 & & -0.113 & & -0.073 &  0.069 \\ 
           & 0.042 & -0.180 & & -0.108 & & -0.287 & -0.123 \\ 
           & 0.040 &  0.180 & & -0.114 & & -0.072 & -0.069 \\ 
           & 0.139 & -0.108 & & -0.107 & & -0.287 &  0.122 \\
\textbf{Tot/f.u.} & 0.091 &  0 & & -0.111 & & -0.180 &  0 \\ 
\mr
\textbf{Co} & -0.542 & -0.004 & & -0.613 & & -0.616 &  0.033 \\ 
            & -0.658 & -0.004 & & -0.611 & & -0.596 & -0.041 \\ 
            & -0.657 &  0.004 & & -0.610 & & -0.615 & -0.033 \\
            & -0.543 &  0.004 & & -0.615 & & -0.596 &  0.042 \\
\textbf{Tot/f.u.} & 0.600 &  0 & & -0.612 & & -0.606 &  0 \\ 
\mr
\textbf{Tot/f.u.} & 0.691 &    & & -0.723 & & -0.786 &    \\ 
\br
\end{tabular}
\end{indented}
\end{table}

Other possibility to obtain the experimental M=0.03$\mu _{B}$ value would
be to have non-collinear magnetism.
UCoGe has four formula units in the unit-cell. To test the possibility of 
non-collinear magnetic order a calculation using the WIEN2k was done but 
with all the atoms in the unit-cell independent, that is, each atom 
could have different values, in particular they could have a different 
magnetic moment directions. WIEN2k is not fully non-collinear, but it can give 
deviations of the magnetic moments from the main magnetization direction 
\cite{Shick}. If as a result of this calculation large deviations are 
obtained then the system should have fully non-collinear magnetism, 
on the other
hand, if only small deviations are obtained then UCoGe would have a main 
magnetization direction with the magnetic moments slightly canted.

With all the atoms independent the symmetry is reduced drastically, the new
space group is $Pm$ (SG \#6). There is no inversion symmetry and complex 
numbers need to be used, this increases fivefold the computation time.
In the new cell the crystal axes are different from the original ones, but 
to avoid confusion the original axes will be used.

Now the most stable magnetization direction is in the $b$-direction, the $a$- 
and $c$-directions are higher in energy by 8.7meV and 0.8meV respectively.

The spin and orbital moments of the case with the main magnetization in the 
$a$-direction are shown in Table 4, the sums are shown in Table 5. The U 
spin moments are relatively large, but they are almost compensated with the
orbital moment. The Co spin moments are smaller, in this case the orbital 
moments are considerably smaller and with the same sign. The canting 
(deviation from the main direction) of the spin terms is very small, the 
largest corresponds to U$_1$=1.6$^\circ$. The canting of the orbital terms are 
larger, the largest is U$_3$=10$^\circ$.

The $c$-direction case shows similar trends, although the U orbital moments are
larger than the spin moments and the system becomes ferromagnetic with the 
U and Co magnetic moments pointing in the same direction. Again, the canting 
of the spin moments is quite small \textless2$^\circ$. For the U atoms the 
canting of the orbital terms is smaller $\sim$5$^\circ$, for the Co atoms the
canting angle is considerably larger Co$_3$=20$^\circ$, but this is due to the 
very small value in the $c$-direction (0.099$\mu_B$). In the $b$-direction 
case, the most stable, again the U orbital moment is larger than the spin term,
in this case, due to symmetry, the magnetization is fully collinear (there is 
no canting).

These results show that in the most stable case there is no canting, while in
the others, not far above, the canting is quite small 
\textless10$^\circ$, with the exception of $c$-direction orbital term 
Co$_3$=20$^\circ$ which, as explained above, is due to the smallness of the 
main term. All these results show that UCoGe does not have a fully 
non-collinear magnetism; all the magnetic moments are aligned, parallel or 
antiparallel, to a main direction with only small deviations.

\begin{figure}[t]
\begin{indented}
\item[]\includegraphics[width=7cm,height=5cm]{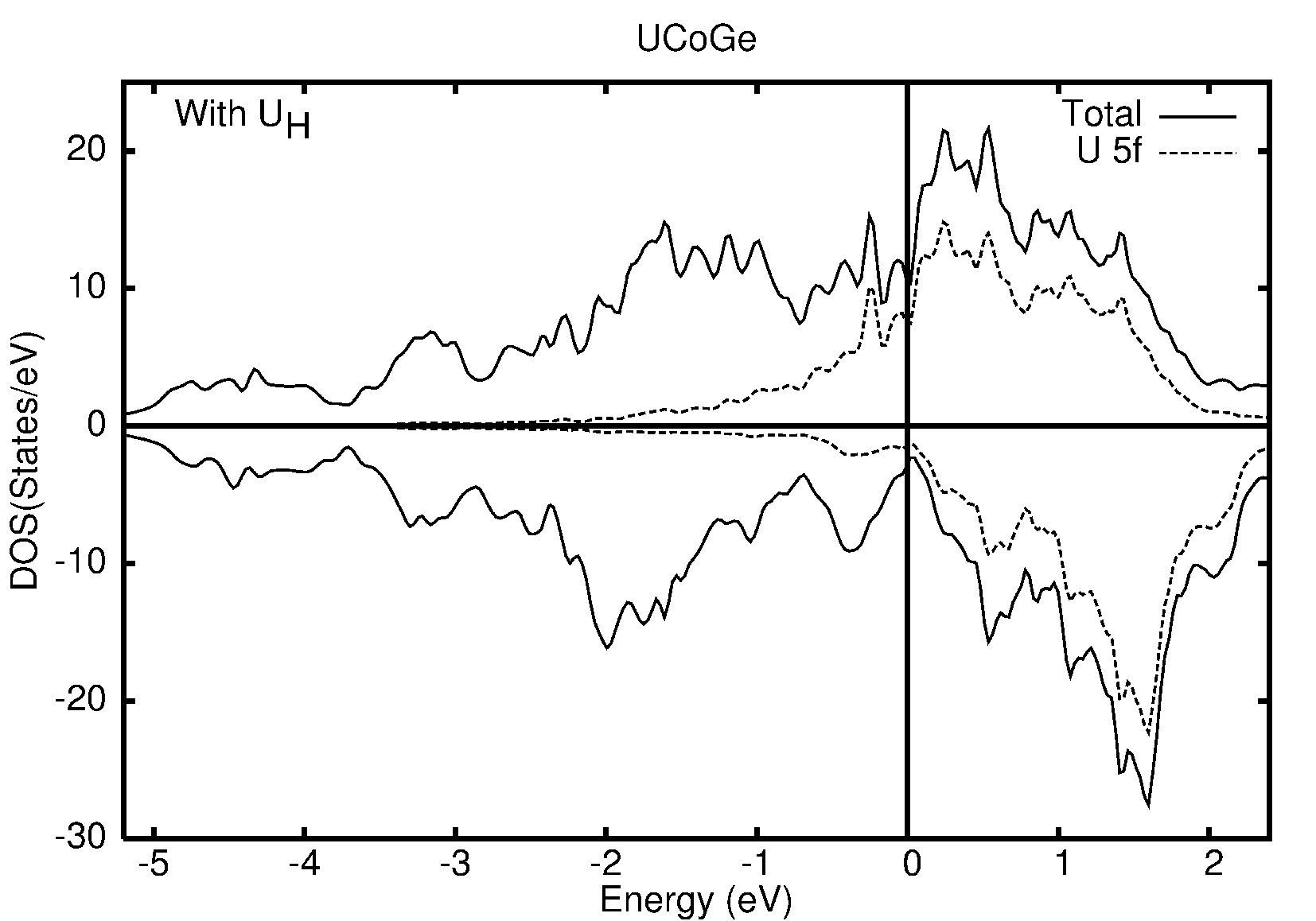}
\caption{\label{Figure4}{\protect\small {DOS for UCoGe ferromagnetic 
calculation with all atoms in the unit cell independent;
up-spin: upward, dn-spin: downward.}}}
\end{indented}
\end{figure}

\begin{figure}[t]
\begin{indented}
\item[]\includegraphics[width=8cm,height=14cm]{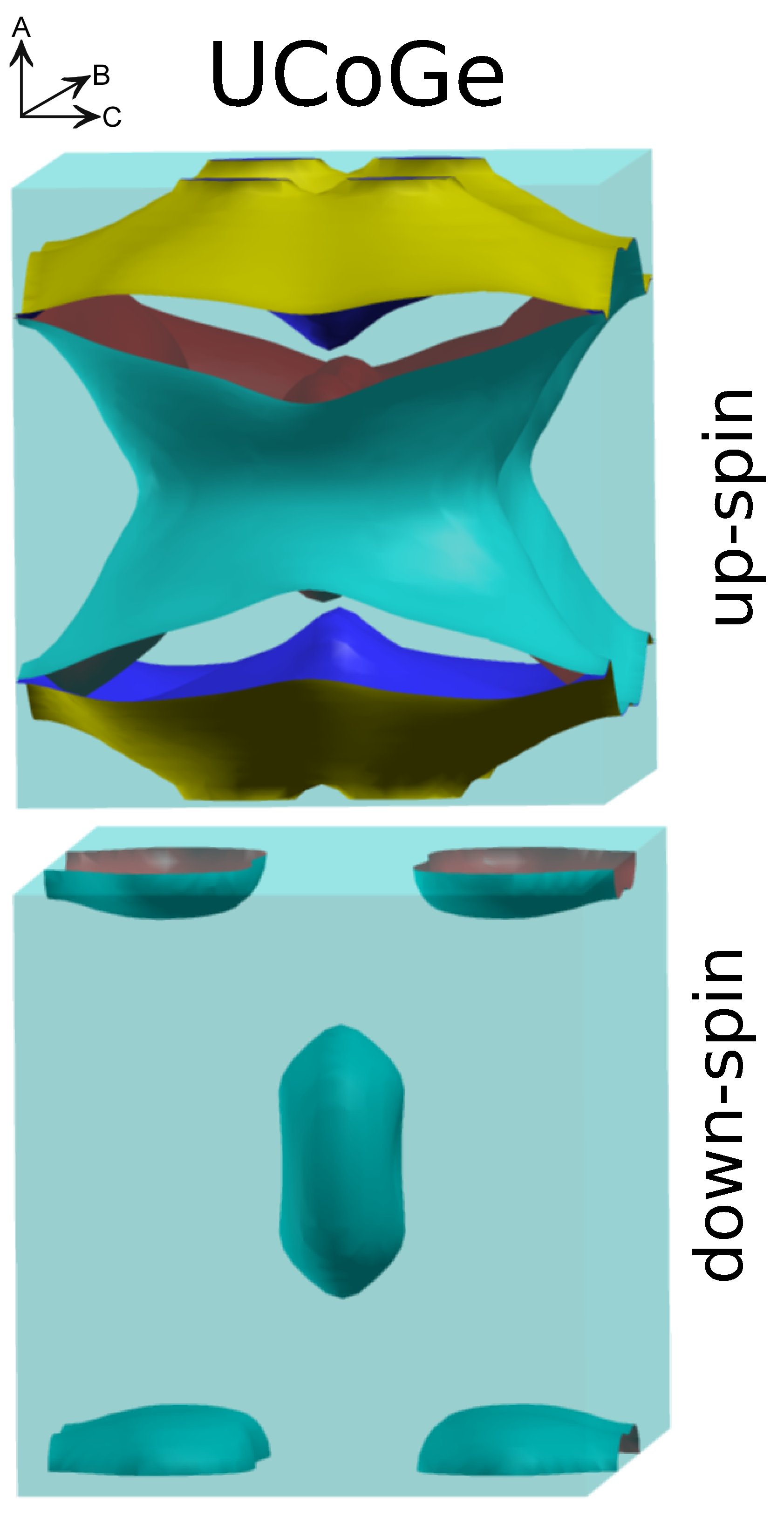}
\caption{\label{figura5}{\protect\small {Fermi surfaces of UCoGe.\\ 
above: up-spin, below: down-spin.}}}
\end{indented}
\end{figure}

The sums of the spin and orbital terms for the three magnetization directions
are shown in Table 5. As it can be seen there are no large differences with the
original case where all the atoms of same kind were equivalent. 
There is an interesting difference though; in the latter case not all the 
moments of each atom are the same, they are split into
two groups, also the canting is in opposite directions, still all the charges 
of equal atoms were found to be the same.

A possible cause of the differentiation of the magnetic values could
be due to the large slope, in the initial case, in the up-spin DOS at $E_F$, 
which is an unstable situation. Given the freedom, in the latter
case, the magnetic moments rearrange and the partial gap is reduced and 
widened (Figure 4).

Fermi surfaces (FS), that are the bands at $E_F$, give information about the 
electrical conductivity; the electrons move perpendicular to the FS. For
parabolic bands it can be easily proven that the conductivity is proportional
to the volume within the FS\cite{MgB2}, this can be applied to FS near the 
band edges, when the bands can be approximated to parabolas.

In UCoGe there are two large up-spin FS (Figure 5), and the material should 
be a good
conductor, in this case it would rule out a phononic mechanism. 
These FS are touching and the possibility of multiple gaps is unlikely 
due to interband scattering. One of the up-spin FS is an undulated plane 
perpendicular to the U-chains, the main conductivity from this FS would be 
along these U-chains (Figure 5, up-spin, top and bottom FS).

For down-spin it is found that $E_F$ is close to the upper band-edge,
DOS has a low value and there are two fairly small FS, all these imply that
the down-spin conductivity is quite low and this material would be close to 
a half-metal.

In the case of URhGe there are large FS in both spin directions, in this 
case they are clearly separated, but still close from each other, 
therefore URhGe seems to be a good conductor and 
probably without multiple gaps. It has a magnetic moment, but in this case 
it is the U-5d orbital that makes the principal contribution to the magnetism.



\section{Conclusion}
Spin polarized electronic structure calculations offer a 
panorama that is essential for the understanding of the coexistence of 
superconductivity and ferromagnetism in UCoGe. These calculations show that
UCoGe is a multiband ferromagnetic superconductor with a magnetic moment 
that is not small which is mainly due to the Co atom. The antiferromagnetic
configuration has a higher energy, also it has been shown that the possibility
of fully non-collinear magnetism is unlikely, the magnetic moments point
towards a definite direction with only small deviations.
The large up-spin Fermi surfaces show that it is a good conductor, probably 
ruling out a phononic mechanism, it has small down-spin Fermi surfaces and
the compound is close to be a half metal.
In the isostructural compound URhGe the magnetic moment is mostly due to the 
U atom.

\ack
We would like to thank Aldo Romero for the valuable suggestions. 
This work was done with support from DGAPA-UNAM under projects PAPIIT IN105207 
and IN108907. P.M. is on sabbatical leave at the IIM-UMSNH with support from 
DGAPA-UNAM.\\

\end{document}